\documentclass[12pt]{article}
\usepackage{epsfig}

\def\beq{\begin{equation}}
\def\eeq#1{\label{#1}\end{equation}}
\def\eeqn{\end{equation}}

\def\beqa{\begin{eqnarray}}
\def\eeqa#1{\label{#1}\end{eqnarray}}
\def\eeqan{\end{eqnarray}}

\let\bar=\overbar

\def\Dslash{\not{\hbox{\kern-4pt $D$}}}
\def\dslash{\not{\hbox{\kern-2pt $\del$}}}

\def\msb{{\bar{\ssstyle M \kern -1pt S}}}

\usepackage{fancyhdr,graphicx}
\def\Title#1{\begin{center} {\Large {\bf #1} } \end{center}}

\begin{document}

\Title{A different nature between the radio AXPs in comparison to the others SGRs/AXPs}

\bigskip\bigskip

%+\addcontentsline{toc}{chapter}{{\it L. Skywalker}}
%+\label{SkywalkerLukeStart}

\begin{raggedright}

{\it J. G. Coelho$^*$ and M. Malheiro\\
Instituto Tecnol\'{o}gico de Aeron\'{a}utica\\
Departamento de Ci\^{e}ncia e Tecnologia Aeroespacial\\
12228-900 Vila das Ac\'{a}cias\\
S\~ao Jos\'{e} dos Campos, SP\\
Brazil\\
{\tt $^*$Email: jaziel@ita.br}}
\bigskip\bigskip
\end{raggedright}

\section{Introduction}

SGRs/AXPs are considered a subclass of pulsars powered by magnetic energy and not by rotation, as normal radio pulsars.
They are understood as strongly magnetized neutron star\cite{Duncan,Thompson}, with large periods
of rotation $P\sim(2-12)$ s, and large spin-down, with typical $\dot{P}\sim(10^{-13}-10^{-10})$ s/s in contrast 
to $\dot{P}\sim10^{-15}$ s for ordinary pulsars. Their
persistent X-ray luminosity, as well as the bursts and flares typical
of these sources, are instead believed to be powered by the decay of
their ultrastrong magnetic field (see Mereghetti 2008 for review~\cite{Mereghetti}).
SGRs/AXPs typically have a larger X-ray luminosity
that can not be explained by their spin-down luminosity ($L_X>\dot{E}_{\rm rot}$), unlike rotation-powered pulsars.
However, the recent discovery of radio-pulsed emission in four of this class of sources, where the spin-down rotational energy lost
$\dot{E}_{\rm rot}$ is larger than the X-ray luminosity $L_X$ during the quiescent state - as in normal pulsars - opens the 
question of the nature of these radio sources in comparison to the others of this class.

According to the fundamental plane for magnetars~\cite{NandaRea},
four over a total of about 20 SGRs/AXPs should have radio-pulsed emission:
XTE J1810-197, 1E 1547.0-5408, PSR J1622-4950, and SGR 1627-41. Basically, the magnetar
radio activity or inactivity can be predicted from the knowledge of the star's
rotational period, its time derivative, and the quiescent X-ray luminosity. However, no radio emission has been
detected in SGR 1627-47 yet because they are unfavorably affected by distance, scattering, or lack of sensitive
observations at the time their pulsed radio emission was possibly
expected to be brighter (see N. Rea et al. 2012 for details~\cite{NandaRea}).

More recently, was reported the discovery of 3.76 s pulsations from the new burst source at the Galactic center
using data obtained with the {\it NuSTAR} Observatory. The SGR J1745−29 is the fourth magnetar detected in radio wavelengths, 
very similar to the others radio SGRs/AXPs.
Also {\it Swift} satellite has observed the sudden turn-on of a new radio source near Sgr A*. This result, combined with the
detection of a short hard X-ray burst from a position consistent with the new radio SGR, suggests that this source is in fact
a new SGR in the Galactic Center. Then, the combination of a magnetar-like burst, periodicity and
spectrum led to the identification of the transient as a likely new magnetar in outburst (see the recent works
about this new SGR in Ref.~\cite{Kennea,Kaya}). 

The radio emission of these sources has several properties that are commom among them, but 
different from the others SGRs/AXPs: loud radio (transient 
radio emission different of the radio pulsar emission), low quiescent X-ray luminosity $L_X$ (decreasing with time) 
that can be explained from the spin-down rotational
energy loss $\dot{E}_{\rm rot}$ of a neutron star, as normal rotation-powered pulsar. 
However, models not invoking neutron stars to describe SGRs/AXPs have also been discussed, in particular
the white dwarf (WD) pulsar model~\cite{MMalheiro,Coelho,Coelho1,Boshkayev}.
As pointed out in Malheiro et al. 2012~\cite{MMalheiro}, the X-ray efficiency $\eta_X=L_X/\dot{E}_{\rm rot}$ for these
radio AXPs, seems to be to small comparing to the others SGRs/AXPs when interpreted as magnetized white dwarfs. However, 
as neutron star pulsars these radio AXPs have $\eta_X\sim(10^{-2}-10^{-1})$, a little bit larger than the values of normal
pulsars (where $\eta_X\sim(10^{-3}-10^{-4})$). This can be understood due to their large magnetic dipole momentum $m\sim10^{32}$ emu,
giving support for their neutron star interpretation. Following the previous work in Ref.~\cite{Coelho3}, we suggest that 
the radio sources are rotation-powered neutron star pulsars $L_X\simeq k\dot{E}_{\rm rot}^n$ ($L_X<\dot{E}_{\rm rot}^{NS}$),
in contrast to the others that are rotation-powered magnetized white dwarfs. In our understanding the large steady X-ray luminosity  
seen for almost all the no-radio SGRs/AXPs, can be explained as coming from a large spin-down 
energy lost of a massive white dwarf with a much large magnetic dipole moment of $10^{34}\leq m\leq10^{36}$
emu consistent with the range observed for isolated and very magnetic WDs (see Coelho \& Malheiro 2012~\cite{Coelho2}
for discussions), indicating a different nature between these sources and the radio SGRs/AXPs.
These radio sources have large magnetic fields and seem to be very similar to the high-B pulsars recently found~\cite{kaspi,Olausen},
as already pointed out in the fundamental plane for radio magnetars~\cite{NandaRea} (see Fig.~1 of this contribution). However, we 
should emphasize that even if the radio SGRs/AXPs are strong magnetized neutron stars, they are not magnetars in the sense 
that their steady luminosity is not originated by the magnetic energy, but from the rotational energy as normal pulsars.

\section{Results and Discussions}
Recently, Rea et al. (2012) try to understand magnetar radio emissions from an phenomenological point of
view. They proposed that magnetars are radio-loud if and only if their quiescent X-ray luminosities are smaller than their rotational
energy loss. 
\begin{figure}
\begin{center}
\psfig{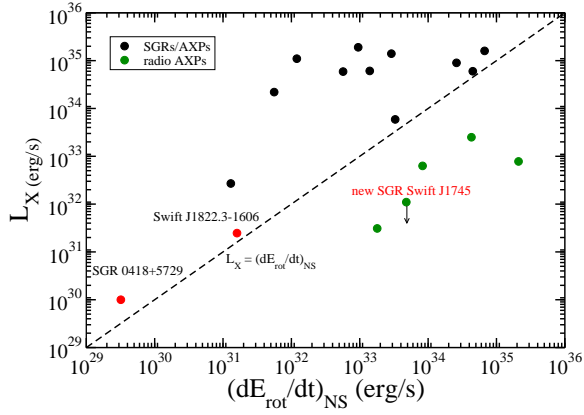}
\end{center}
\caption{X-ray luminosity $L_X$ versus the loss of rotational energy $\dot{E}_{\rm rot}$, 
describing SGRs and AXPs as neutron stars. The red points correspond to recent discoveries of SGR 0418+5729 and 
Swift J1822.3-1606 with low magnetic field. 
}\label{fig1}
\end{figure}
\begin{figure}[hbc!]
\begin{center}
\psfig{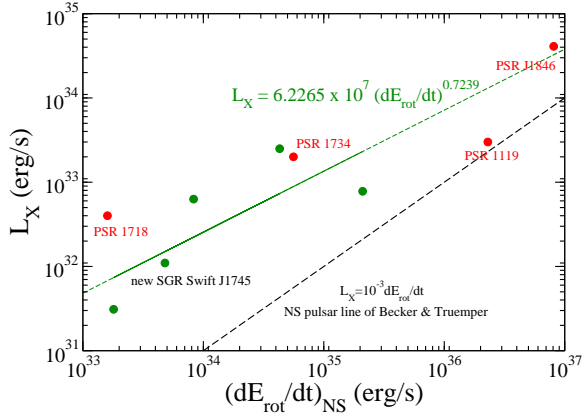}
\end{center}
\caption{X-ray luminosity $L_X$ versus the loss of rotational energy $\dot{E}_{\rm rot}$ for the four
AXPs that show radio-pulsed emission, together with some high-B pulsars. A linear $\log-\log$ relation between $L_X$ 
and $\dot{E}_{\rm rot}$ is found for the radio AXPs, $L_X\propto \dot{E}^{0.7239}$ (green dashed-line),
very similar to the X-ray NS pulsar line of Becker \& Tr$\rm \ddot{u}$mper\cite{Becker} (see McGill SGR/AXP
online catalog - June, 18, 2013 at
www.physics.mcgill.ca/$\sim$pulsar/magnetar/main.html. For pulsar data, see Olausen et al. 2010~\cite{Olausen},
and references therein).
}
\label{fig2}
\end{figure}
\newpage
In Fig.~1, we see a large steady X-ray luminosity (and almost
constant as a function of $\dot{E}_{\rm rot}$) for almost all the no-radio SGRs/AXPs (black circle points - high $\dot{E}_{\rm rot}$
and high $L_X$ - except for the CXO J1647 with a rotational period of 
$P = 10.61$ s, an upper limit of the first time derivative of the rotational period $\dot{P}<4\times10^{-12}$ s/s,
and an X-ray luminosity of $L_X=4.5\times10^{32}$ erg/s), indicating
a different nature among these
sources and the radio SGRs/AXPs (green square points - high $\dot{E}_{\rm rot}$ and low $L_X$). 
These radio sources are in fact ordinary pulsars, 
with their steady X-ray luminosity that can be well explained
within the neutron star model, and with magnetic fields close to the ones of high-B radio pulsars for that
are entirely rotation-powered (see 
Fig.~2). 
We show that the radio SGRs/AXPs obey a linear log-log relation between $L_X$ and $\dot{E}_{\rm rot}$,
$\log L_X= \log(6.2265\times10^7)+0.7239\log{\dot{E}_{\rm rot}}$, very similar to the one
satisfied by X-ray and gamma-ray neutron star pulsars\cite{Becker,Vink,Kargaltsev,Abdo,Arons}, suggesting their neutron star nature.
Furthermore, Fig.~2 shows that almost all the high-B pulsars are also near the line found for the radio AXPs.
In contrast, for almost all the others SGRs/AXPs, $\log L_X$ does not vary too much as
function of $\log \dot{E}_{\rm rot}$, a phenomenology not shared by X-ray neutron star pulsars, suggesting
a different nature for these sources.

\bigskip
The authors acknowledges the financial support of the Brazilian agency CAPES, CNPq
and FAPESP (S\~{a}o Paulo state agency, thematic project $\#$ 2007/03633-3).

\end{document}